\documentclass[12pt]{article}
\usepackage[pdfpagelabels]{hyperref}

\usepackage[usenames,dvipsnames]{color}
\usepackage{graphicx,epsfig,xcolor}
\usepackage{microtype}

\hypersetup{colorlinks=true, linkcolor=violet, urlcolor=blue, citecolor=blue}

\usepackage{cite}
\usepackage[english]{babel}
\usepackage{amssymb,amsfonts,amsmath}
\usepackage{verbatim}
\usepackage{bbm,bbold,bm}
\usepackage{tensor}
\usepackage{slashed}
\usepackage{subcaption}
\usepackage{braket}

\usepackage{xcolor}
\definecolor{darkgreen}{rgb}{0,0.5,0}

\usepackage{comment}
\usepackage{ulem}

\def\ra{\rightarrow}

\def\q{\theta}
\def\a{\alpha}

\def\s{\sigma}
\def\m{\mu}
\def\n{\nu}

\def\diag{\operatorname{diag}}

\def\be{\begin{equation}}
\def\ee{\end{equation}}
\def\ba{\begin{eqnarray}}
\def\ea{\end{eqnarray}}
\def\nb{\nonumber}
\def\p{\partial}

\def\a{\alpha}

\def\g{\gamma}
\def\G{\Gamma}
\def\d{\delta}
\def\D{\Delta}

\def\m{\mu}
\def\n{\nu}
\def\o{\omega}

\def\s{\sigma}
\def\S{\Sigma}

\def\q{\quad}

\def\dd{\mathrm{d}}

\def\ol{\overline}

\def\mf{\mathbf}
\def\mc{\mathcal}

\newcommand{\pr}[1]{\left(#1\right)}
\newcommand{\pq}[1]{\left[#1\right]}
\newcommand{\pg}[1]{\left\{#1\right\}}
\newcommand{\corr}[1]{\left\langle#1\right\rangle}

\usepackage{chngcntr}
\counterwithin*{equation}{section}

\usepackage{pifont}
\usepackage{soul}

\title{\bf Poles-zeros duality in semi-holographic \\Mott insulators}

\author{
Thomas Kögel${}^1$,\,\,
Alessio Caddeo${}^1$,\,\,
Amelie Pitters${}^2$,\\
Francesca Paoletti${}^1$,\,\,
Lorenzo Crippa${}^{3,1}$,\,\,
Giorgio Sangiovanni${}^1$,\\
René Meyer${}^{1,4}$,\,\,
Johanna Erdmenger${}^1$
}

\date{}

\begin{document}
\maketitle 
\vspace{-25pt}
\begin{center}
\begin{minipage}{0.85\textwidth}
\begin{center}
\it{\small 
${}^1$Institute for Theoretical Physics and Astrophysics and Würzburg-Dresden \\ Cluster of Excellence ctd.qmat,  Julius-Maximilians-Universität W\"urzburg \\ Am Hubland, 97074 Würzburg, Germany\\
\vspace{1mm}
${}^2$Bethe Center for Theoretical Physics\\ 
Universität Bonn, 53115 Bonn, Germany}\\
\vspace{1mm}
${}^3$I. Institute of Theoretical Physics, University of Hamburg,\\ Notkestrasse 9, 22607 Hamburg, Germany\\
\vspace{1mm}
${}^4$Shanghai Institute for Mathematics and Interdisciplinary Sciences (SIMIS), Shanghai, 200433, China
\end{center}
\end{minipage}
\end{center}

\begin{center}
\small Email: \texttt{
thomas.koegel@uni-wuerzburg.de,
alessio.caddeo@uni-wuerzburg.de\\ francesca.paoletti@uni-wuerzburg.de}
\end{center}

\vspace{12pt}

\begin{abstract}
\noindent Inspired by the poles-zeros duality of Green's functions that appears in transitions into Mott-insulating phases in  strongly correlated condensed matter systems, we propose
a semi-holographic approach to Mott insulators. In this model, a fundamental fermion is coupled to a large-$N$, strongly interacting sector that generates a self-energy for the fundamental fermion's Green's function. This coupling amounts to a hybridization of the fundamental fermion with a strongly correlated fermionic composite. Within the holographic framework, at large $N$, the Green's function of the composite fermion naturally exhibits a  poles–zeros duality.
Zeros of the Green's function are caused by the poles of the self-energy that correspond to collective many-body excitations of the holographic strongly interacting sector. We calculate the spectral function of the fundamental fermion, from which we characterize the semi-holographic metallic and the Mott-insulating phases. 
In addition to the new physical interpretation of the zeros, our analysis yields a well-defined picture of the poles-zeros duality in terms of the freedom to choose between standard and alternative quantization in the strongly coupled sector.
\end{abstract}

\newpage

\tableofcontents

\vspace{10mm}


\section{Introduction}
\label{sec:intro}

Green's functions and their structure are a central element for the study of particles and excitations in quantum systems, in particular of fermions in condensed matter physics. The two-point Green's function features poles that describe the excitations of the model under consideration.

At weak coupling, where the Landau paradigm holds, these excitations correspond to quasiparticles whose spectrum is in one-to-one correspondence with the free theory. At strong coupling, poles correspond to collective modes of the degrees of freedom involved. Landau Fermi-liquid theory eventually breaks down and single-particle excitations acquire an intrinsic broadening and get shifted to high frequencies. This happens in Mott insulators, i.e. in systems in which strong Coulomb interactions lead to electron localization, even though conventional band theory would predict metallic behavior \cite{Mott_1949}. The insulating gap emerges without long-range order, reflecting the intrinsically many-body nature of the state.

A pole of the Green's function $G(\varepsilon, \mathbf{k})$, with $\varepsilon$ the frequency and $\mathbf{k}$ the momentum, signals an excitation when the energy $\varepsilon$ equals the band dispersion $\varepsilon_n(\mathbf{k})$. The loci in the Brillouin zone where these poles occur at $\varepsilon = 0$ define the Fermi surface.
In addition, in interacting systems, some of the eigenvalues of $G(\varepsilon, \mathbf{k})$ can also vanish at particular points in the Brillouin zone. When the Green's function is a matrix in a multi-flavor space, these points are identified by the vanishing of the determinant of $G(\varepsilon, \mathbf{k})$. 
The set of points where these zeros occur at $\varepsilon = 0$ forms the so-called Luttinger surface \cite{dzyaloshinskii_consequences_2003}.
Initially, such zeros were largely overlooked, but modern theoretical developments have highlighted Green’s function zeros as a hallmark of Mottness, drawing considerable attention within the condensed-matter community \cite{altshuler_luttinger_1998, rosch_breakdown_2007, stanescu_theory_2007, sakai_evolution_2009, sakai_doped_2010, gurarie_single_particle_2011, Gurarie_2011_2,volovik_topology_2012, wang_PRB_2012,Dave_2013,manmana_topological_2012,pudleiner_momentum_2016,Volovik_2018,fabrizio_emergent_2022,fabrizio_PRL_2023,Zhao_2023,setty_symmetry_2023,gavensky_connecting_2023,setty_electronic,Blason_2023,Pasqua_2024,gleis_PRX_2024,Bollmann_2024,Wagner_2024,luaces_PRB_2025,lehmannPRL2025,pangburn_PRB_2025,queiroz2024}. 

\noindent

Green’s-function zeros are particularly relevant in these systems, and although they have been proposed to address open questions in strongly correlated electron materials, determining their exact physical significance remains a major challenge.
For example, recent studies have shown that when topological insulators are driven into a Mott-insulating regime by increasing interactions, new in-gap bands of zeros with nontrivial topology emerge, even though the resulting Hubbard bands are topologically trivial. These in-gap structures are encoded in the zeros of the Green’s function and give rise to corresponding edge zeros. This behavior points to an edge–bulk correspondence for Green’s-function zeros and clarifies how edge poles are converted into edge zeros across the transition \cite{wagner2023, Wagner_2024}.
Taken together, various pieces of evidence indicate that Green's-function zeros carry physical significance as important as the Green's function poles. Understanding them is therefore essential, and it is desirable to develop complementary approaches to fully explore their properties. In this paper we advocate holography as such an approach for better understanding of the role of Green's function zeros in Mott insulators, as well as the observed duality between poles and zeros in Mott transitions.\footnote{See also \cite{Setty:2019ubf,Setty:2021ryz,Grozdanov:2024wgo, Grozdanov:2025ner, Grozdanov:2025ulc} for other studies relating poles and zeros.}

Since Mott insulators are intrinsically non-perturbative states of matter, analytic approaches are generally impractical, and weak-coupling expansions and conventional renormalization-group schemes fail to capture their essential physics. A small number of tractable limits include the atomic limit, the  one-dimensional Hubbard model ($d=1$) exactly solvable via the Bethe ansatz, or the opposite limit of infinite dimensions ($d=\infty$) \cite{metzner1989}, where a numerically exact solution is provided by dynamical mean-field theory (DMFT) \cite{GeorgesRMP}. Beyond these developments, despite impressive numerical progress in recent years, our theoretical understanding remains limited \cite{Gull2013, LeBlanc2015,Schaefer2021}. This scarcity of reliable non-perturbative methods motivates the exploration of alternative frameworks capable of addressing strong correlations beyond these special cases.

A powerful tool for studying strongly interacting systems is the holographic correspondence \cite{Maldacena:1997re, Witten:1998qj, Gubser:1998bc}, which maps the strongly coupled dynamics of a quantum field theory to a gravitational theory in a spacetime with an additional dimension. This framework has also proven useful in the context of condensed matter systems \cite{Liu:2009dm,vcubrovic2009string,Faulkner:2009wj} (see, e.g., \cite{Hartnoll:2018xxg, Hartnoll:2009sz, Zaanen:2015oix,Ammon:2015wua} for reviews). 
In particular, in the context of Mott physics, it was shown in \cite{Edalati:2010ww, Edalati:2010ge} (see also \cite{Vanacore:2014hka, Vanacore:2015poa, Seo:2018hrc}) that the Green’s function of a massless bulk fermion with a non-minimal\footnote{As opposed to the minimal coupling to the gauge field by the gauge covariant derivative $D_\mu=\partial_\mu-iqA_\mu$.} Pauli dipole coupling in an anti–de Sitter (AdS) Reissner–Nordström black hole background captures the appearance of Green's function zeros in the spectral gap, depending on the strength and sign of the dipole coupling. First studies of a duality between poles and zeros in the holographic context were presented in \cite{Alsup:2014uca, Vanacore:2014hka}. Moreover, it was recently shown by Sin and collaborators in \cite{Ghorai:2024nxs} that a non-minimal scalar coupling $\bar\Psi \Psi F^2$ for the bulk fermion also gives rise to properties of Mott physics and is arguably preferable to the dipole coupling case, as it gives rise to a spectral function with a more symmetric and hard Mott gap.

On the other hand, despite exhibiting features of Mott phenomenology, the holographic Green’s function of a probe bulk fermion does not satisfy the single-particle sum rules required for canonical (anti-)commutation relations, thereby complicating direct comparison with conventional condensed matter systems. 
While sum rules for composite fermionic operators in strongly coupled conformal field theories were derived in \cite{Gulotta:2010cu}, these are different in nature from the canonical ones for fundamental fermions.
A way to circumvent this problem  and still retain the advantages of holography is provided by \textit{semi-holography} \cite{Faulkner:2010tq} (see also\cite{Hartnoll:2009ns, Nickel:2010pr, Faulkner:2010jy}). This approach introduces a fundamental field coupled to a strongly interacting sector with a holographic gravity dual. From the holographic perspective, this amounts to promoting the sources of the composite operator in the strongly coupled sector to dynamical fields. The Green's function of such a fundamental field then features a self-energy that is determined by the holographic theory, and satisfies single-particle sum rules \cite{Gursoy:2011gz}.

In this paper, we take a step forward in the analysis and understanding of the poles-zeros duality by establishing it in a semi-holographic model, thereby providing new insight into its working mechanisms  at least in the strongly correlated setting. We propose a semi-holographic setup in which a fundamental fermion (an electron) is coupled to a composite fermionic operator of a strongly interacting sector. The composite fermion has a holographic dual in terms of a massless bulk fermion in an AdS Reissner–Nordström black hole background. The bulk fermion action includes the non-minimal coupling of \cite{Ghorai:2024nxs}. Combining the semi-holographic approach with this coupling has two appealing features: the Green’s function of the semi-holographic fermion inherits characteristic features of Mott phenomenology through its coupling to the strongly interacting sector, while also satisfying the single-particle sum rules\footnote{Further works relating semi-holography to Mott physics with a different approach and focus are \cite{Doucot:2020fvy,Samanta:2022myh}.}. The benefit of our semi-holographic approach to the poles-zeros duality is an explanation of the appearance of Green's function zeros as collective excitations in the strongly coupled sector.

Let us summarize our main results. First, following \cite{Alsup:2014uca} we show that the model's self-energy given, in our setup, by the Green's function of the composite fermionic operator, exhibits an exact poles-zeros duality. That is, the self-energy is mapped into its inverse by changing the sign of the bulk non-minimal scalar coupling parameter. As a result, the poles of the self-energy become zeros if the sign of the control parameter is changed. Moreover, we find that the poles of the self-energy cause the appearance of zeros of the fundamental fermion's Green's function. As explained above, this is a distinctive feature of Mott physics. We thus see that, in our semi-holographic setup, the emergence of zeros is tied to excitations of the strongly coupled sector, and is therefore inherently a many-body effect associated with its large-$N$ nature. We further show that the exact poles–zeros duality of the self-energy induces an approximate poles–zeros duality in the Green’s function of the fundamental fermion at low frequencies and momenta,  in closer spirit to the duality observed in condensed matter studies. 
 
We perform a detailed numerical study of the fundamental fermion Green’s function as a function of the non-minimal scalar coupling of the bulk fermion. In this analysis, we observe a crossover between a conducting and a Mott-insulating behavior. More precisely, at large positive coupling the system is in a state characterized by sharpened peaks and simple poles of the single-particle Green's function. We refer to this phase as \textit{semi-holographic metal}. As the coupling is reduced toward zero, the system enters a regime characterized by broad spectral features and ultimately the loss of quasiparticle coherence. For sufficiently large negative coupling, a spectral gap opens, which we identify as a Mott gap when a zero of the Green’s function lies within it. This numerical investigation also confirms our analytical derivation of the approximate poles-zeros duality. We find that the linearly dispersed in-gap zeros of the single-particle correlator in the Mott regime are, with high accuracy, mapped into poles in the semi-holographic metallic phase.

It is interesting to compare how the poles-zeros duality emerges in the condensed-matter context via many-body solutions of Hubbard models, as opposed to our semi-holographic setup. In the former case, the poles-zeros duality emerges from a direct analysis of the Green's function and the self-energy in the Mott insulator \cite{wagner2023,lehmannPRL2025, setty_symmetry_2023, setty_electronic}.
Yet, this approach comes with challenges as it involves a thermodynamic phase transition separating the Mott phase from the Fermi-liquid metal. Thus, the relation between the momentum dispersion of the zeros with the free band structure is not a simple mapping. In the semi-holographic setup, the poles-zeros duality emerges from the possibility to choose between two different ways of coupling the fundamental to the composite operator of the strongly coupled sector. In the holographic context, these two ways are referred to as \textit{standard} and  \textit{alternative quantization} and correspond to two different choices of boundary conditions imposed on the bulk field dual to the composite operator of the strongly coupled sector \cite{Klebanov:1999tb}. In our setup, switching between the two quantizations is equivalent to changing the sign of the non-minimal scalar coupling. Describing poles and zeros in this way can help to better understand their common nature as  excitations \cite{fabrizio_emergent_2022,fabrizio_PRL_2023,Wagner_2024}.

The paper is organized as follows. In Section \ref{sec:holosetup}, we present the semi-holographic setup review the bulk model. We study the strongly coupled sector, and derive the poles-zeros duality for the self-energy and for the Green's function. 
In Section \ref{sec:Results}, we present our model’s phenomenology by a detailed numerical analysis. 
In Section \ref{sec:discuss}, we provide a final discussion and outlook.



\section{Semi-holographic model}
\label{sec:holosetup}
Here we present the semi-holographic setup, developed and expanded in \cite{Faulkner:2010tq, Hartnoll:2009ns, Nickel:2010pr, Faulkner:2010jy}, that we use to model the crossover between a Mott-insulating state and a holographic metal. The setup consists of a continuum quantum field theory in $2+1$ dimensions and features a Dirac fermion coupled to a strongly coupled holographic sector. By renormalizing the free propagator of the fundamental fermion the strongly coupled sector will thus be responsible for the Mott physics we observe.
After reviewing some basic aspects of semi-holography, we focus on the strongly coupled sector, which, via the AdS/CFT correspondence \cite{Maldacena:1997re, Witten:1998qj, Gubser:1998bc}, is modeled by a classical gravitational theory living in $3+1$ dimensions.

\subsection{Semi-holography}
\label{sec:semiholography}
The idea of semi-holography is to study, within a quantum field theory framework, the dynamics of fields that are coupled to a strongly interacting sector whose behavior can be analyzed through holographic methods. For the type of semi-holographic construction to be introduced in the following paragraphs, this will generally involve a field with canonical (anti)commutation relations linearly coupled to a strongly interacting sector. The upshot of this construction is that the renormalization of the propagator due to strong interactions may be calculated directly via the AdS/CFT correspondence.

We consider a Dirac fermion $\chi$ coupled to a fermionic operator $\mc{O}$. The latter is, in general, a single-trace composite operator built out of the fields of the strongly coupled sector. The action of the entire system takes the form
\be
\label{eq:CFTAction}
S[\chi, \overline \chi]  = - i  \int d^{3} x \overline \chi \g^{\m} \pr{ \p_{\m} - i q A_{\m}} \chi  + S_{\text{strong}}^{(\mc{O})} 
+ i g \int d^{3} x \pr{\ol{\mc{O}} \chi + \ol{\chi} \mc{O} } \ .
\ee
Here, $\g^{\m}$ are gamma matrices in $(2+1)$ dimensions, which satisfy the Clifford algebra $\pg{\g^{\m}, \g^{\n}} = 2 \eta^{\m \n} \mathbb{1}$, with $\m=0,1,2$. We use the representation
\be
\label{eq:boundarygammamatrices}
\g^{0} = i \s^{2} \ , \q \q \q  
\g^{1} = \s^{1} \ , \q \q \q  
\g^{2} = \s^{3} \ , 
\ee
where $\s^{1,2,3}$ are Pauli matrices. The field $A_{\m} = \m \d^{0}_{\m}$ introduces the chemical potential $\m$, $q$ is the charge of $\chi$ and $g$ is a free coupling parameter. We work in the approximation in which $\chi$ does not backreact on the strongly coupled sector, which will be discussed in more detail in Section \ref{sec:stronglysector}. For the moment, it suffices to note that this sector exhibits $\sim N^{2}$ degrees of freedom, and we work in the large-$N$ limit. In this regime, the strongly coupled sector can be described holographically by a classical gravitational theory in $3+1$ dimensions, in which the fermionic operator $\mc{O}$ is dual to a fermionic field $\Psi$ and $\chi$ plays the role of a source for the operator $\mc{O}$. 

The main quantity of our interest is the retarded two-point function of $\chi$,
\be
\label{def:greensfunctionchi}
G_{R}(t,\vec x) =   i \theta (t) \corr{ \{\chi (t, \vec x), \chi^{\dagger} (0,\vec 0)\} }.
\ee
Another important object is the retarded Green's function of the composite operator $\mathcal{O}$, 
\be
\label{def:greensfunctioncomposite}
\q \mc G_{R}(t,\vec x) =   i \theta (t) \corr{ \{ \mc{O} (t, \vec x), \mc{O}^{\dagger} (0,\vec 0)\} } \ .
\ee
With these conventions, the imaginary part of $G_{R}$ is positive definite \cite{Iqbal:2009fd}.
As we will show in Sections \ref{sec:stronglysector} and \ref{sec:poleszeroessemiholo}, in the large-$N$ limit, the strongly coupled sector can be integrated out to give
\be
\label{eq:resultsingleparticleGreenfunction}
G_{R}  = \pr{ G_{0}^{-1} + g^{2} \g^{0} \mc G_{R} \g^{0}}^{-1} \ , 
\ee
where $G_{0}$ is the $\chi$ propagator at $g=0$. As a result, the strongly coupled sector provides a self-energy contribution
\be
\label{eq:SelfEnergyHolographic}
\Sigma_R = g^{2} \g^{0} \mc{G}_{R} \g^{0} \ .
\ee 

Here, it is already important to emphasize that it is precisely the large-$N$ limit which, as we discuss in more detail in the following sections, allows the self-energy correction to the single-particle Green’s function to be represented by a simple two-point function. In this sense, the holographic sector acts as a bath for the semi-holographic fermion $\chi$, and the self-energy should be understood as encoding the hybridization between the electron and a strongly interacting medium. As laid out in the next subsection, our model contains a free parameter, $\eta$, which controls the behavior of the self-energy. Crucially, in Section \ref{sec:polezeros}, we will show that poles (zeros) of the retarded Green's function of the holographic sector $\mathcal{G}_R$ at positive values of $\eta$ are mapped exactly into zeros (poles) at the corresponding negative values of $\eta$. We will then argue that, under appropriate assumptions, this will give rise to an approximate duality between poles and zeros for the single particle propagator $G_R$. Consequently, varying $\eta$ drives the system from a incoherent metal to either a Mott-insulating or coherent metallic phase.

\subsection{Strongly coupled sector}
\label{sec:stronglysector}
The strongly coupled sector of our model is a $(2+1)$-dimensional gauge theory with a large number $N^{2}$ of gluon degrees of freedom plus a fermionic flavor sector with $\sim N$ degrees of freedom, to which the fermionic operator $\mc{O}$ of Section \ref{sec:semiholography} belongs. Taking a bottom-up holographic approach, we define the gauge theory through a $(3+1)$-dimensional bulk theory featuring a metric $g_{MN}$ describing the gluon sector, a gauge field $A_{M}$ introducing a finite chemical potential, and a Dirac fermion $\Psi$ that captures the physics of the flavor sector.
The fermion $\Psi$ is holographically dual to the fermionic operator $\mc{O}$. The action is
\be
\label{eq:completeholoaction}
S_{\text{Holo}} = \frac{1}{2 \kappa_{4}^{2}} \int d^{4} x \sqrt{-g} \pq{ R + \frac{6}{L^{2}} - L^{2} F^{2}  - \frac{i}{L N} \ol \Psi \pr{ \overset{\leftrightarrow}{\slashed{D}} - m -  \eta L^{3} F^{2}  } \Psi  } \ .
\ee 
Here, $L$ is the AdS radius, $F^{2} \equiv F^{MN} F_{MN}$ and
\be
\slashed{D} = e^{M}_{a} \G^{a} \pr{\p_{M} + \frac{1}{4} \o_{M}^{bc} \G_{bc} - i q A_{M}} \ , \q \q \q \G_{ab} = \frac{1}{2} \pq{\G_a,\G_{b}} \ ,  
\ee
where $e^{M}_{a}$ is the vielbein, $\o_{M}^{ab}$ the spin connection, $A_{M}$ the gauge field whose field strength is $F_{MN}$, and $q$ the charge of the Dirac fermion. Moreover, $\G^{a}$ are Dirac matrices in four dimensions, satisfying\footnote{A set of matrices that satisfy this Clifford algebra in $3+1$ dimensions is
\label{eq:setofgamma}
\begin{align}
\label{eq:chiralbasisGamma}
    \Gamma^{3} &= \begin{pmatrix} \mathbb{1} & 0 \\ 0 & - \mathbb{1} \end{pmatrix} \ , 
    \!&\!
    \Gamma^{\m} &= \begin{pmatrix} 0 & \g^{\m} \\ \g^{\m} & 0  \end{pmatrix}  \ ,
\end{align}
where $\g^{\m}$ satisfy the Clifford algebra in $2+1$ dimensions.
}
\be
\pg{\G^{a}, \G^{b}} = 2 \eta^{ab} \mathbb{1} \ , \q \q \q \eta^{ab} = \diag \pr{-1, 1, 1, 1} \ .
\ee
and $\overset{\leftrightarrow}{\slashed{D}}$ just means that for the partial-derivative term we symmetrize the left and right derivative (including a $1/2$ factor).\footnote{We adopt uppercase Latin letters for bulk coordinates, lowercase Latin letters from the beginning of the alphabet for tangent space indices, and lowercase Latin indices from the middle of the alphabet for spatial boundary indices. We keep using lowercase Greek letters for boundary spacetime coordinates.} 
In the action (\ref{eq:completeholoaction}), we included factors of $L$ so that the coupling parameter $\eta$ is defined as a dimensionless quantity.
 
We work in the probe approximation in which we neglect the backreaction of $\Psi$ on the background, which is given by the AdS Reissner-Nordström black-hole solution
\begin{subequations}
\label{eq:RNsolution}
\be
ds^{2} = \frac{L^{2}}{z^{2}} \pr{ - f(z) dt^{2} + f(z)^{-1} dz^{2} + dx^{i} dx^{i} } \ , \q \q A = \m \pr{1 -  \frac{z}{z_{h}}} \dd t \ ,
\ee
where $\m$ is the chemical potential and the blackening factor $f(z)$ is
\be
f(z) = 1 - \pr{1+  \m^{2} z_{h}^{2}  } \pr{\frac{z}{z_{h}}}^{3} +   \m^{2} z_{h}^{2}   \pr{\frac{z}{z_{h}}}^{4} \ .
\ee
\end{subequations}
In these coordinates, the horizon is at $z=z_{h}$ and the conformal boundary at $z=0$. In some of the computations, it will be useful to introduce a cut-off on the holographic direction. In that case, the boundary will be at $z=z_{b}$, with $z_{b}$ very small.
The temperature is given by
\be
\label{eq:deftemperature}
T = - \frac{f'(z_{h})}{4 \pi z_{h}} =  \frac{3-  \m^{2} z_{h}^{2} }{4 \pi z_{h}}  \ .
\ee
Due to the underlying scaling symmetry, only the ratio $T/\m$ is physically meaningful.

The physics relevant to Mott phenomenology originates from the flavor sector
\be
\label{eq:fermionholoaction}
S_{f} = \frac{i  \a_{f}}{L^{3}} \int d^{4} x \sqrt{-g}   \ol \Psi \pr{ \overset{\leftrightarrow}{\slashed{D}}  - m -  \eta  L^{3}F^{2} } \Psi    \ ,
\ee 
where $\a_{f} = -L^{2}/2 \kappa_{4} N$. Notice that typically $L^{2}/2\kappa_{4} \sim N^{2}$, thus $\a_{f} \sim N$.  
The Dirac fermion is coupled to the background gauge field $A_{M}$ not only through the covariant derivative, but also through the last term in (\ref{eq:fermionholoaction}). As we will see, it is precisely the value of the parameter $\eta$ that determines whether the system is Mott insulating or in a conducting phase. The idea of introducing a non-minimal coupling goes back to \cite{Edalati:2010ww, Edalati:2010ge} (see also \cite{Vanacore:2014hka, Vanacore:2015poa, Seo:2018hrc}), where a dipole-like coupling $ e^{M}_{a} e^{N}_{b} F_{MN} \, \bar\Psi \Gamma^{ab} \Psi $ was shown to give rise to Mott physics. The scalar coupling in (\ref{eq:fermionholoaction}) was proposed in \cite{Ghorai:2024nxs}, where it was found that the corresponding spectral function exhibits a symmetric Mott gap, in contrast to the dipole coupling. Moreover, \cite{Ghorai:2024nxs} argues that the scalar coupling can be more naturally mapped to the interaction term of the Hubbard model.

The equations of motion for the Dirac spinor are
\be
\label{eq:diraceq}
\pr{ \slashed{D}  - m -   \eta L^{3}  F^{2} } \Psi  = 0 \ .
\ee
Close to the boundary $z=0$, the general solution is
\ba
\Psi^{\text{bdy}}( z) &\sim&  z^{3/2-mL} \pq{  \begin{pmatrix} 0 \\ \chi_{-}
\end{pmatrix} + z \begin{pmatrix} 0 \\ \chi_{-}^{\text{sub}} 
\end{pmatrix} +  O(z^{2})} \nb \\
&+&  z^{3/2+ mL} \pq{ \begin{pmatrix} \chi_{+} \\ 0
\end{pmatrix}  + z \begin{pmatrix} \chi_{+}^{\text{sub}} \\ 0
\end{pmatrix} + O(z^{2}) }\ ,
\ea
where $\chi_{\pm}$ and $\chi_{\pm}^{\text{sub}}$ are Dirac spinors in $2+1$ dimensions and we suppressed the dependence on the boundary coordinates. The parameter $\eta$ does not appear in the near-boundary solutions. 
It is convenient to define
\be
\label{def:psipmpsimp}
\Psi_{\pm} = \frac{1}{2}\pr{\mf{1} \pm \G^{3}} \Psi \ , \q \q \q \G^{3} \Psi_{\pm} = \pm \Psi_{\pm} \ .
\ee
Thus, 
\begin{subequations}
\label{eq:boundarysolutions}
\ba
\Psi^{\text{bdy}}_{-}(z) &=&   z^{3/2-mL}   \pq{  \begin{pmatrix} 0 \\ \chi_{-}
\end{pmatrix} + z \begin{pmatrix} 0 \\ \chi_{-}^{\text{sub}} 
\end{pmatrix} +  O(z^{2})}  \ , \\
\Psi^{\text{bdy}}_{+}(z) &=&  z^{3/2+ mL} \pq{ \begin{pmatrix} \chi_{+} \\ 0
\end{pmatrix}  + z \begin{pmatrix} \chi_{+}^{\text{sub}} \\ 0
\end{pmatrix} + O(z^{2}) } \ .
\ea
\end{subequations}
The solutions get exchanged under the transformation $m \ra -m$, hence we can consider $m \geq 0 $ without loss of generality. For generic value of the mass, one imposes Dirichlet boundary conditions on $\Psi_{+}$. This is what is called standard quantization, where $\chi_{+}$ is interpreted as the source for a (fermionic) operator $\mc{O}_{\text{st}}$ with conformal dimension $\D_{\text{st}} = 3/2 + m L$, whereas $\chi_{-}$ gives its expectation value. When $|mL|< 1/2$, one is allowed to instead fix the value of $\Psi_{-}$. This choice is called alternative quantization, and gives rise to a boundary theory where $\chi_{-}$ is interpreted as the source of an operator $\mc{O}_{\text{alt}}$ with conformal dimension $\D_{\text{alt}} = 3/2- mL$, whereas $\chi_{+}$ gives its expectation value \cite{Klebanov:1999tb, Henneaux:1998ch}. In this paper, we set $m=0$, so our fermionic operator $\mc{O}$ has conformal dimension $\D = 3/2$ for both choices of boundary conditions.

We are interested in computing the retarded Green's function $\mc{G}_{R}$ defined in (\ref{def:greensfunctioncomposite}). The holographic prescription \cite{Son:2002sd, Iqbal:2008by, Iqbal:2009fd, Gubser:2008sz} instructs us to solve the equations of motion (\ref{eq:diraceq}) fixing $\chi_{+}$ and imposing in-falling boundary conditions at the horizon at $z=z_{h}$. In this way, we obtain a relation of the form
\be
\label{eq:relationbetweenchi}
\chi_{-} = \mc{S} \cdot \chi_{+} \ , 
\ee
where the dot indicates a convolution if we work in coordinate space and a multiplication if we work in momentum space. Notice that $\mc{S}$, not to be confused with the action $S$, is a two-by-two matrix. The one-point function of $\mc{O}$ is given by the conjugate momentum with respect to the holographic direction through\footnote{We define $\ol \Pi_{+} = \p \mc{L}/\p \p_{z} \ol \Psi_{+}$, which has four components. However, when evaluating $\lim_{z\ra 0} z^{-3/2} \ol \Pi_{+}$, only  its last two components contribute. With a slight abuse of notation, we continue to denote these last two components by $\ol \Pi_{+}$.} \cite{Iqbal:2009fd}
\be
\corr{\mc{O}}_{\chi_{+}} = i  \pr{ \lim_{z \ra 0} z^{-3/2} \ol \Pi_{+}}_{\text{on-shell}} =  \a_{f} \, \mc{S} \cdot \chi_{+} \ .
\ee
From linear response theory, we know
\be
 \corr{ \mc{O}}_{\chi_{+}} = i \mc{G}_{R} \g^{0} \chi_{+} \ , 
\ee
and therefore we obtain
\be
\label{eq:retardedstrongcorrelator}
\mc{G}_{R} =  i \a_{f}   \mc{S} \g^{0} \ .
\ee
The same result can be, equivalently, obtained \cite{Gursoy:2011gz} by introducing the boundary action
\be
\label{eq:Sbdy}
S_{\text{bdy}} =  \frac{i \a_{f}}{L^{3}} \int_{z=z_{b}} d^{3} x \sqrt{-h}  \,  \ol \Psi_{+} \Psi_{-}   \ ,
\ee
where $h$ is the determinant of the metric induced on the boundary at $z=z_{b}$. Evaluating it on-shell, we find
\be
S^{\text{on-shell}}_{\text{bdy}} =   i \a_{f}   \int_{z=z_{b}} d^{3} x d^{3} y \,   \ol \chi_{+} (x) \mc{S}(x-y) \chi_{+} (y)  \ .
\ee
Using that $S_{f}$ vanishes on-shell, from the holographic prescription
\be
\corr{\exp \pr{- \int d^{3} x \pr{ \ol \chi_{+} \mc{O} + \ol{\mc{O}} \chi_{+} }}}_{\text{CFT}} = e^{i S^{\text{on-shell}}_{\text{bdy}}[\chi_{+}, \ol \chi_{+}]}  \ , 
\ee
we indeed retrieve (\ref{eq:retardedstrongcorrelator}). 

The latter approach is particularly suitable for embedding the strongly coupled sector in the semi-holographic framework \cite{Gursoy:2011gz}. Indeed, we can make the source $\chi_{+}$ dynamical by adding a kinetic term on the boundary,
\be
\label{eq:Skin}
S_{\text{kin}} = - \frac{i}{g^{2} L^{3}}  \int_{z=z_{b}} d^{3} x\; \sqrt{-hg_{zz}}\; \ol \Psi_{+} \;e^{\m}_{\underline{a}} \G^{\underline{a}}\; \pr{\p_{\m} - i q A_{\m}} \Psi_{+} \ , 
\ee
where $g$ is a free parameter, $\underline{a} = 0,1,2$ and the index $\m$ runs over the boundary directions. Recalling (\ref{eq:boundarysolutions}), this becomes
\be
S_{\text{kin}} = - \frac{i}{g^{2} }  \int  d^{3} x \;   \ol \chi_{+}\; \g^{\m }  \pr{\p_{\m} - i q A_{\m}} \chi_{+} \ .
\ee
We thus define $S_{\text{semi}} = S_{\text{kin}} + S^{\text{on-shell}}_{\text{bdy}}$ and introduce the generating functional 
\be
Z[J] = \int D \ol \chi_{+} D \chi_{+} e^{i S_{\text{semi}}[\chi_{+}, \ol \chi_{+}] - \int d^{3} x \pr{\ol J \chi_{+} + \ol \chi_{+} J}} \ .
\ee
From this, we can compute the (retarded) Green's function for $\chi_{+}$.\footnote{More precisely, the two-point function for $\chi_{+}$ is found through
\be
\corr{\chi_{+} (t, \vec x) \chi_{+}^{\dagger} (0, \vec 0)} = - \pr{\frac{\d^{2} Z[J]}{\d J 
(0, \vec 0) \d \ol J (t, \vec x)} }_{J=0}\g^{0} \ .
\ee} Identifying $\chi = \chi_{+}/g$, its Green's function then reads
\be
G_{R} = \pr{i \g^{0} \g^{\m} \pr{ \p_{\m} - i q A_{\m}} - i \a_{f}g^{2} \g^{0} \mc{S}}^{-1} \ .
\ee
Recalling (\ref{eq:retardedstrongcorrelator}) and defining $G_{0}^{-1} = i \g^{0} \g^{\m} \pr{\p_{\m} - i q A_{\m}}$, this can be rewritten as
\be
\label{eq:semiholocorrelator}
G_{R}  = \pr{ G_{0}^{-1} + g^{2} \g^{0} \mc G_{R} \g^{0}}^{-1} \ ,
\ee
which is the result anticipated in (\ref{eq:resultsingleparticleGreenfunction}). From (\ref{eq:retardedstrongcorrelator}), we see that the correlator $\mc{G}_{R}$ scales as $\a_{f} \sim N$. Consequently, if the coupling $g$ parameter is independent of $N$, the self-energy scales as $\mc{O}(N)$.
\subsection{Poles-zeros duality of the self-energy}
\label{sec:polezeros}
The Green's function $\mc{G}_{R}$ is characterised by a poles-zeros duality: once we fix $|\eta|$, the poles (zeros) at $+\eta$ are mirrored by zeros (poles) at $-\eta$. Such a duality was found in \cite{Alsup:2014uca} for the dipole-like coupling $ e^{M}_{a} e^{N}_{b} F_{MN} \, \bar\Psi \Gamma^{ab} \Psi $. Here, we follow the same steps for the scalar coupling.

From the holographic point of view, the origin of the poles-zeros duality is clear \cite{Vanacore:2014hka}. As we commented in Section \ref{sec:stronglysector}, when the mass of the bulk fermion $\Psi$ vanishes, $m=0$, there is no physical difference between the gauge theories in standard and alternative quantization, as they are connected by the relabelling $\Psi_{\pm} \ra  \pm \Psi_{\mp}$, where $\Psi_{\pm}$ are components of $\Psi$ defined in (\ref{def:psipmpsimp}). However, one can easily check that while the radial derivative term in (\ref{eq:fermionholoaction}) is even under such a relabelling, the scalar coupling term is odd. This means that changing quantization is equivalent to changing the sign of $\eta$. On the other hand, from (\ref{eq:relationbetweenchi}), we see that the relabelling maps $\mc{S} \ra  -  \mc{S}^{-1}$, which amounts to $\mc{G}_{R} \ra  \a_{f}^{2} \g^{0} \mc{G}_{R}^{-1} \g^{0}$. As a result, if $\mc{G}_{R}$ displays a pole (zero) at a given positive value of $\eta$, it must also display a zero (pole) at $-\eta$.

Let us show the poles-zeros duality in a more explicit way by rewriting the Dirac equations (\ref{eq:diraceq}). It is convenient to work with the spinors $\psi_{\pm}$ defined through
\be
\Psi  = \pr{-g g^{zz}}^{-1/4} \psi \ , \q \q \q \psi =  \begin{pmatrix} \psi_{+} \\ \psi_{-} 
\end{pmatrix} \ .
\ee
We also take advantage of the rotational symmetry to align the spatial momentum $\vec{k}$ along the $x^{1}$ direction. In Fourier space, we have 
\begin{subequations}
\label{eq:Diracequationstwobytwo}
\ba
\pr{  f(z)   \p_{z} -   \eta L^{4} \frac{\sqrt{f(z)}}{z} F^{2} } \psi_{+} &=&   i   \pq{  \pr{\o + q  A_{0}} \g^{0}  -  \sqrt{f(z)} k \g^{1}} \psi_{-}   \ , \\
- \pr{ f(z)   \p_{z} +   \eta L^{4} \frac{\sqrt{f(z)}}{z} F^{2} } \psi_{-} &=&   i   \pq{  \pr{\o + q  A_{0}} \g^{0}  -  \sqrt{f(z)} k \g^{1}} \psi_{+}   \ . \q \q 
\ea
\end{subequations}
Recalling (\ref{eq:boundarygammamatrices}), we see that since $\g^{0}$ and $\g^{1}$ have only off-diagonal elements, the equations of motion imply that the matrix $\mc{S}$ has only non-diagonal components,  
\be
\label{eq:formofS}
\mc{S} = \begin{pmatrix} 0 & \mc{S}_{+} \\ \mc{S}_{-} & 0 \end{pmatrix}   \ , \q \q \q \psi_{-} = \mc{S} \psi_{+} \ .
\ee
Thus, the Green's function $\mc{G}_{R} = i \a_{f}  \mc{S} \g^{0}$, in our basis where $\g^{0} = i \s^{2}$ will be
\be
\label{eq:holographicGreenfunction}
\mc{G}_{R} = - i \a_{f}  \lim _{z \ra z_{b}} \begin{pmatrix}
\mc{S}_{+} & 0 \\ 0 &  - \mc{S}_{-}
\end{pmatrix} \ , \q \q \q \det \mc{G}_{R} =  \a_{f}^{2} \lim _{z \ra z_{b}}  \mc{S}_{+} \mc{S}_{-} \ .
\ee
Plugging $\psi_{-} = \mc{S} \psi_{+} $ into  (\ref{eq:Diracequationstwobytwo}), we obtain two decoupled non-linear equations:
\ba
    \label{eq:FlowEqS}
  f(z) \p_{z}  \mc{S}_{\pm}  \mp  i \pr{ \omega + q A_{0}} 
   \left(\mc{S}_{\pm}^2-1\right) +   \sqrt{f(z)}
   \left[\frac{2 \eta  L^{4} F^{2}  }{z}\mc{S}_{\pm} - i k \pr{
   \mc{S}_{\pm}^{2} + 1}\right] = 0  \ .  \q 
\ea
It is easy to see that performing the transformation
\be
\mc{S}_{\pm} \ra  \frac{1}{\mc{S}_{\pm}} \ , \q \q k \ra -k \ ,  \q \q \q \eta \ra - \eta
\ee
the equations remain the same. As a result, we obtain the relation
\be
\label{eq:ExactPZD}
\mc{G}_{R} (\o, k, \eta) = -  \a_{f}^{2} \mc{G}_{R}^{-1} (\o, -k, -\eta) \ ,
\ee
which establishes the poles-zeros duality: poles (zeros) at a fixed value $+|\eta|$ correspond to zeros (poles) at $-|\eta|$. In our semi-holographic setup, this is interpreted as a poles-zeros duality of the self-energy $\Sigma_{R}$ of the fundamental boundary fermion $\chi$. 
\begin{figure}[t]
  \centering
  \includegraphics[width=1\linewidth]{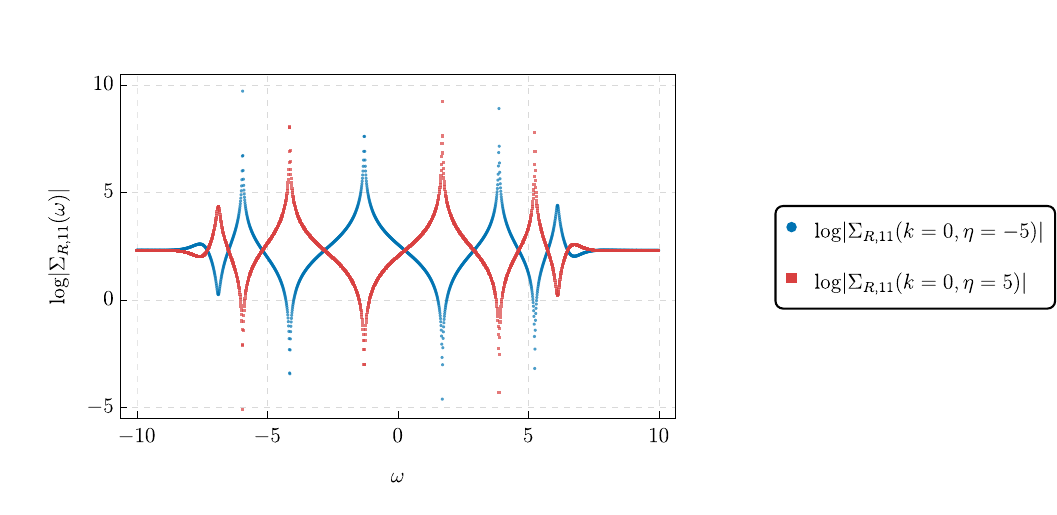}
  \caption{Plot of the logarithm of $ \vert \S_{R,11}(\omega,k=0,\eta=5)\vert $ (blue) and $ \vert  \S_{R,11}(\omega,k=0,\eta=-5)\vert $ (red) over frequency $\omega$ at $k=0$. As predicted by Eq. (\ref{eq:dualSE}), poles (zeros) at $\eta = 5$ are mapped into zeros (poles) at $\eta=-5$. The self-energy, related to the holographic correlator $\mc{G}_R$ by Eq. (\ref{eq:SelfEnergyHolographic}), goes to a constant at large $\omega$, consistently with the discussion in \cite{Gulotta:2010cu}.}
\label{fig:AbsDetSelfEnergypm5}
\end{figure}
Indeed, recalling (\ref{eq:SelfEnergyHolographic}), from (\ref{eq:ExactPZD}), we get
\be
\label{eq:dualSE}
\Sigma_{R}(\omega, k, \eta) = - \a_{f}^{2} g^{4} \Sigma_{R}^{-1}(\omega, -k, -\eta) \ .
\ee

A graphical representation of the relation (\ref{eq:dualSE}) on the level of the component $\Sigma_{R,11}$ at zero momentum is shown in Fig. \ref{fig:AbsDetSelfEnergypm5} where we plot the logarithm of the absolute value to visually amplify the poles and zeros respectively. Poles show up as positive and zeros as negative divergencies. For any pole (zero) at $\eta = 5$, there corresponds a zero (pole) at $\eta = -5$. Moreover, at large frequencies, $\Sigma_R$ tends to a non-zero constant. This behavior is expected, since in our setup $\Sigma_R$ is determined by the Green's function $\mc{G}_R$ of a composite operator $\mc{O}$ in the strongly coupled sector. When the mass of the bulk fermion $\Psi$ vanishes, $\mc{G}_R$ approaches a constant at large frequencies \cite{Gulotta:2010cu}, consistent with an emergent conformal symmetry in the ultraviolet, where the chemical potential - responsible for breaking scale invariance - becomes negligible. As a consequence, $\mc{G}_R$ does not satisfy single-particle sum rules \cite{Gulotta:2010cu}. This limitation motivates the semi-holographic setup, which instead focuses on the Green's function $G_R$, behaving as $G_R \sim -1/\omega$ at large frequencies (see Eq. (\ref{eq:semiholocorrelator})) and therefore satisfying single-particle sum rules \cite{Gursoy:2011gz}.

The numerical results presented in Section \ref{sec:Results} are obtained by solving Eq. (\ref{eq:FlowEqS}). One can see that the component $\pr{\mc{G}_{R}}_{11}$ at momentum $k$ equals the component $\pr{\mc{G}_{R}}_{22}$ at momentum $-k$. The solution giving rise to the retarded Green's function are found imposing in-falling boundary conditions, which in terms of the flow variables $\mc{S}_{\pm}$ read
\begin{subequations}
\label{eq:bdycond}
\be
\lim_{z \ra z_{h}} \mc{S}_{\pm} = \mp 1
\ee
for any value of the temperature $T$ and non-vanishing frequency $\omega$. In the case of vanishing frequency, $\omega=0$, and vanishing temperature $T=0$, the boundary conditions are
\be
\lim_{z \ra z_{h}} \mc{S}_{\pm} = i \frac{6  \sqrt{2} + \sqrt{72 \eta ^2  + 2  z_{h}^{2 } k^2 - q^2} }{\sqrt{2} z_{h} k \pm q} \ .
\ee
\end{subequations}
All the numerical results presented in Section \ref{sec:Results} are at zero temperature, which, recalling (\ref{eq:deftemperature}), corresponds to $\mu= \sqrt{3}/z_h$. 
In numerical computations, it is also convenient to set $z_h=1$. Accordingly, quantities such as the frequency and momentum are measured in units of $\mu/\sqrt{3}$.

\subsection{Poles and zeros in the semi-holographic setup}
\label{sec:poleszeroessemiholo}
In this section, we want to discuss the implications of the exact poles-zeros duality of $\Sigma_R$ in our semi-holographic setup. In particular, we will see that the exact poles-zeros duality of the self-energy entails an approximate poles-zeros duality for the Green's function $G_{R}$ of the elementary fermion $\chi$ (see Eqs. (\ref{def:greensfunctionchi}) and (\ref{def:greensfunctioncomposite}) for the definition of our Green's functions.)

Let us start by reminding that, in the large-$N$ limit, the AdS/CFT correspondence in standard quantization can be framed, up to possible counterterms, in the equation
\begin{align}
     W_{\mathrm{CFT}}[J,\bar{J}]=S_{\mathrm{grav}}^{\text{on-shell}}[J,\bar{J}] \ ,
\end{align}
where $W_{\mathrm{CFT}}[J,\bar{J}]$ is the generating functional of connected Green's functions, such that $Z[J,\bar{J}]=\exp(iW_{\mathrm{CFT}}[J,\bar{J}])$. As explained in Section \ref{sec:stronglysector}, in our setup $J$ is the boundary value of the bulk mode which we take to be the source for the operator to which it couples. Thus, we may formally rewrite (\ref{eq:CFTAction}) in terms of the integration over the strongly coupled sector as
\begin{align}
    e^{iW_{\text{CFT}}[g\chi,g\bar{\chi}]}=\int\mathcal{D}\{ \Phi \}_{\text{CFT}}\;\exp\left( i S_{\text{strong}}^{(\mc{O})} 
+ i g \int d^{3} x \pr{\ol{\mc{O}} \chi + \ol{\chi} \mc{O} }\right)\ .
\end{align}
Hence, the total effective action for the fundamental fermion $\chi$ can be written as
\begin{align}
    S_{\text{eff}}[\chi,\bar \chi]=S_{\text{kin}}[\chi,\bar \chi]+W_{\text{CFT}}[g\chi,g\bar{\chi}] \ ,
\end{align}
where
\begin{align}
   S_{\text{kin}}[\chi,\bar \chi]=- i  \int d^{3} x \overline \chi \g^{\m} \pr{ \p_{\m} - i q A_{\m}} \chi \ .
\end{align}
Since the generating functional for the single-trace sector $W_{\text{CFT}}$ truncates at the level of the two-point functions in the large-$N$ limit, we again obtain the Dyson relation written in Eq. (\ref{eq:semiholocorrelator}).

We have yet to make a statement on the quantization prescription used in the dual holographic theory. In principle, we can eschew this discussion entirely by noting that for a strongly coupled sector described by the theory (\ref{eq:completeholoaction}) we can achieve this flipping the sign of the bulk coupling $\eta$ as discussed in Section \ref{sec:polezeros}. This operation then leads to an exchange of the poles and zeros of the self-energy. From the field theory point of view this is not obvious, so we recall another relation between the generating functionals of standard and alternative quantization found by \cite{Klebanov:1999tb}, which we now call $W_{\mathrm{CFT}}^{\text{std}}[J,\bar{J}]$ and $W_{\mathrm{CFT}}^{\text{alt}}[J,\bar{J}]$ respectively. They are related by a Legendre transformation of the boundary action,
\begin{align}
    W_{\mathrm{CFT}}^{\text{alt}}[\alpha,\bar{\alpha}]=S_{\mathrm{grav}}^{\text{alt, on-shell}}[\alpha,\bar{\alpha}]=S_{\mathrm{grav}}^{\text{on-shell}}[J,\bar{J}]-\int d^dx \pr{ \bar{\alpha}J+\bar{J}\alpha} \ ,
\end{align}
where $\alpha(x)=\frac{\delta W_{\mathrm{CFT}}^{\text{std}}[J,\bar{J}]}{\delta \bar J(x)}$ and analogously for the conjugate. The Legendre transformation of $W[J,\bar{J}]$ with respect to the expectation value gives the one-particle irreducible action $\Gamma[\alpha,\bar{\alpha}]$. Here, we have used the general relation
\begin{align}
    \frac{\delta^2 W[J,\bar J]}{\delta J(x)\delta \bar J(y)}= \frac{\delta \bar\alpha(x)}{\delta \bar J(y)} = \left(\frac{\delta \bar J(y)}{\delta\bar \alpha(x)}\right)^{-1}= - \left( \frac{\delta^2\Gamma[\alpha,\bar\alpha]}{\delta \alpha(x)\delta \bar \alpha(y)} \right)^{-1} \ .
\end{align}
This equation entails a relation between the Green's functions in standard and alternative quantization that, omitting proportionality factors, reads
\begin{align}
    G^{\text{alt}}_R \sim -(G^{\text{std}}_R)^{-1} \ .
\end{align}
This implies the relation (\ref{eq:SelfEnergyHolographic}) according to which the poles of the self-energy are mapped onto zeros and vice versa.

As already mentioned, our main focus is on the poles and, in particular, the zeros of the single-particle correlator (\ref{eq:semiholocorrelator}) as the bulk coupling parameter $\eta$ is varied. The zeros of (\ref{eq:semiholocorrelator}) are determined by the poles of the self-energy, and their dispersion defines the Luttinger surface in the Mott-gapped phase. The poles of the individual components of $G_R$ are, on the other hand, given by solutions to
\be
\label{eq:dyson}
    G_0^{-1}(\omega,k)+\Sigma_{R}(\omega,k)=0 \ .
\ee
Because the bare propagator $G_0$ appears in the Schwinger-Dyson relation, the exact poles--zeros duality satisfied by the self-energy $\Sigma_R$ cannot hold exactly for the single-particle Green's function $G_R$. 
Nonetheless, an approximate duality can survive in limited regions of $\omega$ and $k$ (as in \cite{wagner2023}).

Indeed, if the coupling parameter $g$ is $\mathcal{O}(N^0)$, the self-energy scales as $\mc{O}(N)$ and therefore dominates the Schwinger-Dyson equation. 

We can see this more clearly in the following way. Take $\omega_p$ to be a simple pole of $\mathrm{det}\;G_R$ at some given momentum $k$. Thus, the matrix $M=G_0^{-1}+\Sigma$ has a one-dimensional kernel with corresponding left and right null vectors at $\omega_p$,
\begin{align}
    Mv=0 \ ,  \q \q \q u^\dagger M=0 \ .
\end{align}
We normalize them as $u^\dagger v=1$ and define
\begin{align}
    m(\omega)=u^\dagger M v \ .
\end{align}
Since we are interested in locating the pole of the determinant close to the zero of the self-energy, we take its position $\omega_p$ to be of the form $\omega_p = \omega_* + \delta\omega$. Expanding $m(\omega)$ around $\omega_p$, to first order in $\delta\omega$, we obtain
\begin{align}
    0=m(\omega_p)=m(\omega_*)+\delta\omega \partial_\omega m(\omega_*)+\mathcal{O}(\delta\omega^2) \ .
\end{align}
Solving for $\delta\omega$ yields
\begin{align}
    \delta\omega=-\frac{m(\omega_*)}{\partial_\omega m(\omega_*)}=-\frac{u^\dagger M(\omega_*)v}{u^\dagger \partial_\omega M(\omega_*)v}=-\frac{u^\dagger G_0^{-1}(\omega_*)v}{u^\dagger\left( \partial_\omega G_0^{-1}(\omega_*)+\partial_\omega\Sigma(\omega_*)\right)v} \ .
\end{align}
In our basis, $G_R$ is diagonal, and we can simplify this to
\begin{align}
    \delta\omega=-\frac{\omega_*\pm k}{1+\partial_\omega\Sigma_{11/22}(\omega_*,k)} \ ,
\end{align}
depending on which entry of the matrix propagator $G_R$ features a pole.
Since the denominator of this expression scales with $\alpha_fg^2\sim N$, the shift $\delta\omega$ goes as $\delta\omega\sim 1/N$. Thus, the greater $N$ is, the more precise the poles-zeros duality becomes. 

We can also understand why the duality breaks down at large frequencies for fixed, large but finite $N$. Since $\Sigma(\omega,k)$ asymptotes to a constant at large $\omega$, while $G_0^{-1}$ grows linearly, there is a regime of frequencies and momenta in which $\delta\omega$ starts to increase. More precisely, for large $\omega$ the self-energy approaches $\alpha_f g^2$, and the shift eventually becomes of order $\mathcal{O}(1)$. Hence, for sufficiently large values of $N$ we find that, within the region 
\begin{align}
|\omega|,\;|k| &< | \alpha_f | g^2 \ , 
\end{align}
the exact poles–zeros mapping of the self-energy $\Sigma_R$ induces an approximate poles-zeros correspondence for the single-particle Green's function $G_R$, which becomes progressively less accurate outside this window.

Let us conclude this section with some comments on the physical interpretation of the mechanism behind the appearance of the zeros of the Green's function $G_{R}$ in our semi-holographic setup. 

As discussed above, the zeros of $G_R$ are determined by the poles of the self-energy $\Sigma_R$. The pole surface of $\Sigma_{R}(\omega,k)$ is associated with the Fermi surface of the composite operator $\mathcal{O}(\omega,k)$ in the strongly coupled sector, which is in turn governed by the infrared CFT to which the system flows \cite{Faulkner:2009wj}. As a consequence, in a Mott-insulating phase, the low-energy properties of the electron are controlled by collective excitations.\footnote{Here, collective in the sense that an excitation of $\mathcal{O}$, and the associated poles, correspond by the nature of holographic duality to excitations involving $\mathcal{O}(N)$ degrees of freedom. It is therefore an intrinsically many-body effect.} More concretely, this means that properties such as the location and the sharpness of the Luttinger surface or the Hubbard bands are controlled by data such as the infrared scaling dimension of $\mc{O}$. 

It is now apparent that, in order to obtain a zero close to the Fermi level (or, equivalently, a pole of the propagator of $\mathcal{O}$) as well as a poles–zeros duality for $G_R$, the large-$N$ limit is required in our setup. Only in this limit does $\mathcal{O}$ behave as a generalized free field \cite{Duetsch:2002hc}, exhibiting well-defined particle-like properties. Thus, in our setup, the emergence of Mottness is directly tied to a composite operator acquiring quasi-particle-like behavior. More broadly, the semi-holographic model captures an important aspect of Mott phenomenology: the gapping of the semi-holographic fermion is accompanied by a transfer of spectral weight to collective excitations of the strongly coupled sector.


\section{Numerical results and phenomenology}
\label{sec:Results}

We now turn to our numerical results. We begin by examining the poles--zeros duality and then show that switching on the bulk coupling drives the system from a incoherent metal phase into either a Mott insulating or a more coherent metallic phase, depending on the sign and magnitude of the coupling $\eta$.

The relevant quantities are obtained by numerically solving Eq.~(\ref{eq:FlowEqS}) with the appropriate in-falling boundary conditions \eqref{eq:bdycond} at the black hole horizon. Once $\mathcal{S}_\pm$ are determined, the holographic Green's function $\mathcal{G}_R$ is constructed using Eq.~(\ref{eq:holographicGreenfunction}), and the corresponding self-energy is then obtained from
Eq.~(\ref{eq:SelfEnergyHolographic}). The single-particle Green's function $G_R$ is subsequently computed via Eq.~(\ref{eq:resultsingleparticleGreenfunction}). Note that both the self-energy $\Sigma_R$ and the single-particle Green's function $G_R$ depend on the parameters $\alpha_f$ and $g$ only through the combination $\alpha_f g^2$. In what follows, all numerical results are presented for $\alpha_f g^2 = -10$, with the bulk fermion charge fixed to $q = 1$.

A key feature of the Green’s function $G_R$ is the presence or absence of a spectral gap, i.e. a range of energies around zero frequency where the spectral density vanishes. The gap edges and gap sizes reported in this section have been extracted, focusing for simplicity, on the spectral density at $k=0$ only
\begin{align}    
\rho(\omega,k) =\frac{1}{\pi} \mathrm{Tr}\,\mathrm{Im}\, G_R(\omega,k)\,.
\end{align}
Numerically, we identify the gap primarily by imposing the criterion
$\rho(\omega, k=0) < 10^{-2}$. However, since the determination of the gap
inevitably involves some degree of arbitrariness, we also evaluate it
using the more stringent cutoff $\rho(\omega, k=0) < 10^{-5}$. Taken together, these two cutoffs provide
reasonable estimates for lower- and upper-bounds of the gap values across a wide
interval of~$\eta$. Unless otherwise stated, all gaps and band edges
displayed in the figures throughout the remainder of this section are
determined using the more lenient cutoff.

We restrict our analysis to the zero-temperature case $T = 0$, which, according to Eq. (\ref{eq:deftemperature}), is achieved by setting $\mu = \sqrt{3}/z_h$. With $z_h = 1$, all frequencies and momenta are expressed in units of $\mu/\sqrt{3}$.

\subsection{Poles-zeros duality}

\begin{figure}[t!]
\begin{subfigure}{.5\textwidth}
  \centering
  \includegraphics[width=.91\linewidth]{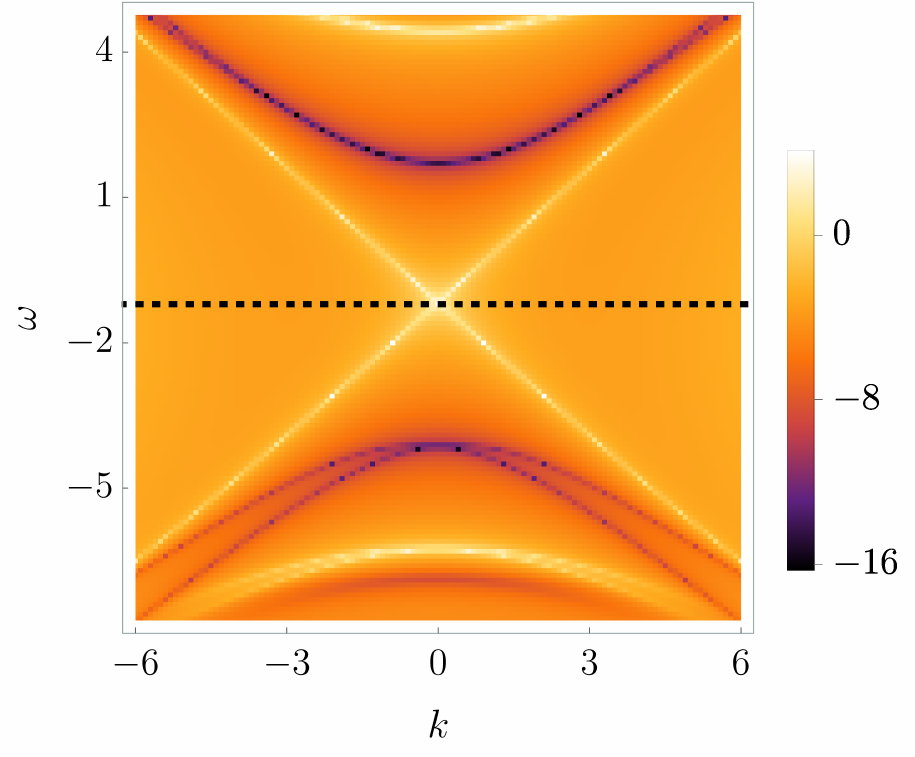}
  \caption{Plot of $\mathrm{log}\vert \det G_R(\omega,k,\eta=5) \vert$}
  \label{fig:Poles}
\end{subfigure}%
\begin{subfigure}{.5\textwidth}
  \centering
  \includegraphics[width=.9\linewidth]{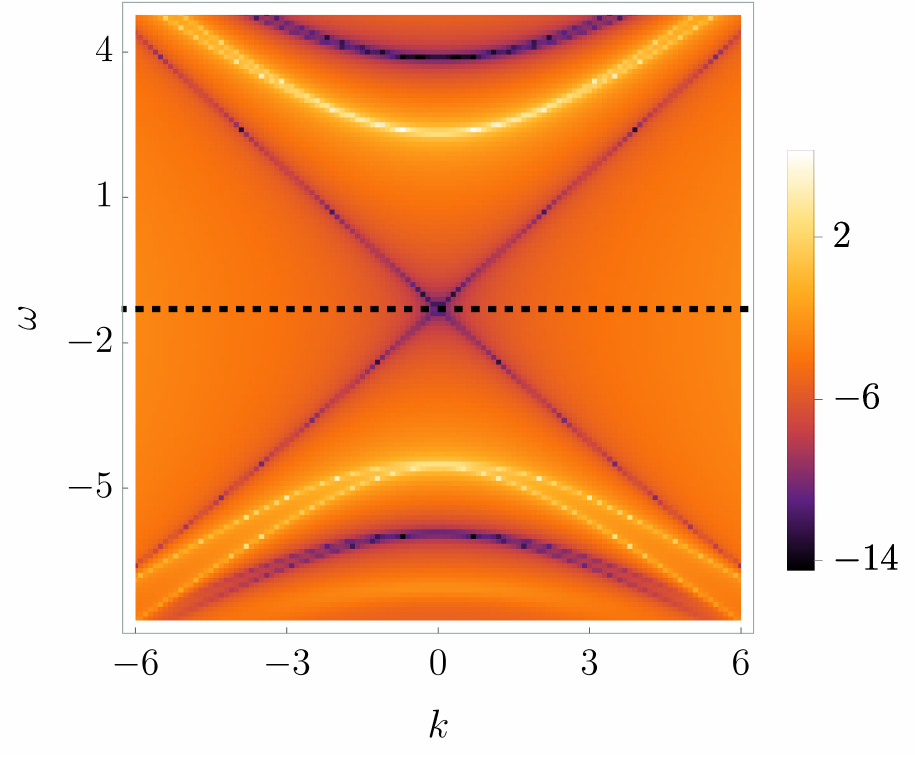}
  \caption{Plot of $\mathrm{log}\vert \det G_R(\omega,k,\eta=-5) \vert$}
  \label{fig:Zeroes}
\end{subfigure}

\caption{Comparison of $\log \vert \det G_R(\omega,k,\eta) \vert$ for opposite-sign values of the bulk coupling, $\eta = 5$ (metallic, in panel (a)) and $\eta = -5$ (Mott insulating, in panel (b)). Bright yellow regions indicate peaks in the quantity, while dark purple regions correspond to zeros. The horizontal dashed lines represent the Fermi level which is $\omega_F \approx -1.2$ for the metallic phase (left) while it is fixed in the middle of the gap, i.e. at $\omega_F \approx -1.3$ in the Mott insulator (right).
The poles and zeros of $\vert \det G_R(\omega,k,\eta) \vert$ exhibit an approximate one-to-one correspondence over the displayed range of momenta and frequencies.}
\label{fig:Poles&Zeroes}
\end{figure}

In order to analyze the dispersion of poles and zeros, we inspect the quantity $\log \left| \det G_R(\omega,k,\eta) \right|$. The reason is twofold. On the one hand, zeros for multi-component/multi-flavor systems are encoded in the eigenvalues \cite{gurarie_single_particle_2011, wagner2023,volovik_quantumVacuum} rather than in the diagonal elements of the Green's function matrix. On the other hand, the logarithm makes it easy to visualize both poles and zeros in the same plot.
In Fig.~\ref{fig:Poles&Zeroes}, poles appear in yellow while zeros are shown in purple, according to our color gradient. Interestingly, we observe that the linearly dispersing poles of $\vert \det G_R(\omega,k,\eta) \vert$ close to the Fermi level for $\eta = 5$ (Fig.~\ref{fig:Poles}) qualitatively map onto the zeros of the same quantity at opposite coupling $\eta = -5$ (Fig.~\ref{fig:Zeroes}).

\begin{figure}
  \centering
  \includegraphics[width=1\linewidth]{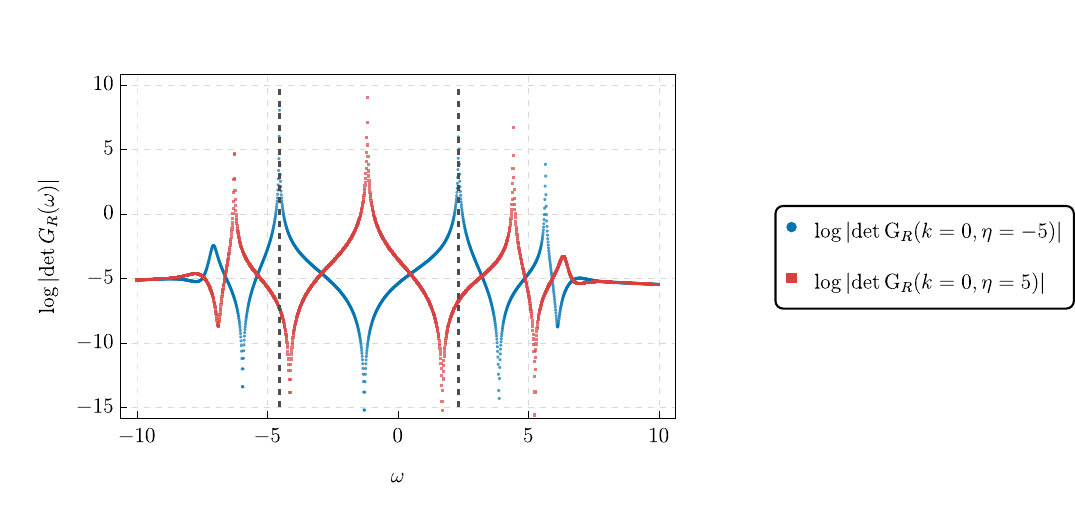}
  \caption{Plot of $\mathrm{log}\vert \det G_R(\omega,k,\eta) \vert$ at $k=0$ over $\o$ deep in the Mott and metallic phase at $\eta=5$ and $\eta=-5$ respectively. The zeros and the poles can be seen to match to a high level of accuracy. The gap in the Mott phase, determined from the edges of the spectral density $\rho(\omega,k)$ at $k=0$, is here of size $\Delta_{\mathrm{gap}}=6.82$ in units of $\mu/\sqrt{3}$, and is marked by the grey dashed vertical lines. The isolated, low-energy zero can be seen to sit inside the gap.}
\label{fig:PolesZeroesDetk0}
\end{figure}

A cut at $k=0$ of $\log \vert \det G_R(\omega,k,\eta) \vert$ as a function of $\omega$ for $\eta = -5$ and $\eta = 5$ is shown in Fig.~\ref{fig:PolesZeroesDetk0} for better clarity. From this plot, the correspondence between poles and zeros becomes particularly clear: poles appear as positive divergences, while zeros as negative divergences. The close matching between these features at opposite values of the coupling is thus immediately evident. This should be compared to the exact matching of poles and zeros for $\eta=\pm 5$ at $k=0$ in Fig. \ref{fig:AbsDetSelfEnergypm5}. The location of the zero of $\det G_R$ remains unchanged in comparison to the divergencies of $\Sigma_{R,11}$, while the poles experience a minor shift, as anticipated in the previous section.

\begin{figure}
\begin{subfigure}{.5\textwidth}
  \centering
  \includegraphics[width=1.0\linewidth]{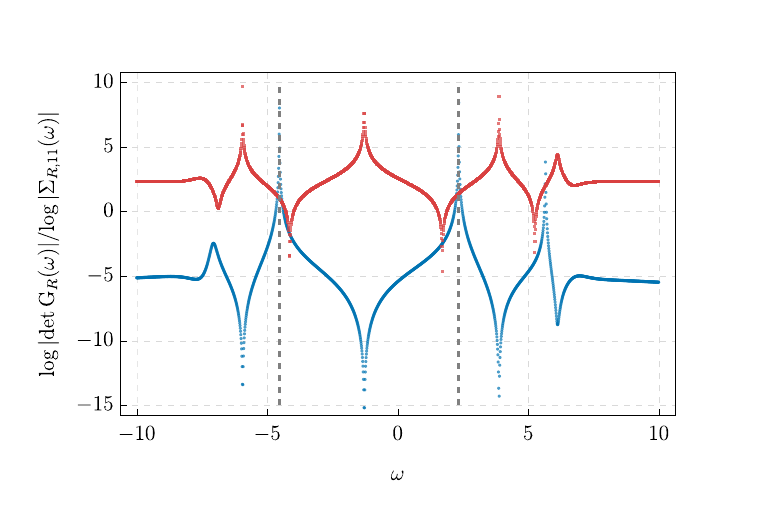}
  \caption{$\eta=-5$}
  \label{fig:DetGSigmaMott}
\end{subfigure}%
\begin{subfigure}{.5\textwidth}
  \centering
  \includegraphics[width=1.0\linewidth]{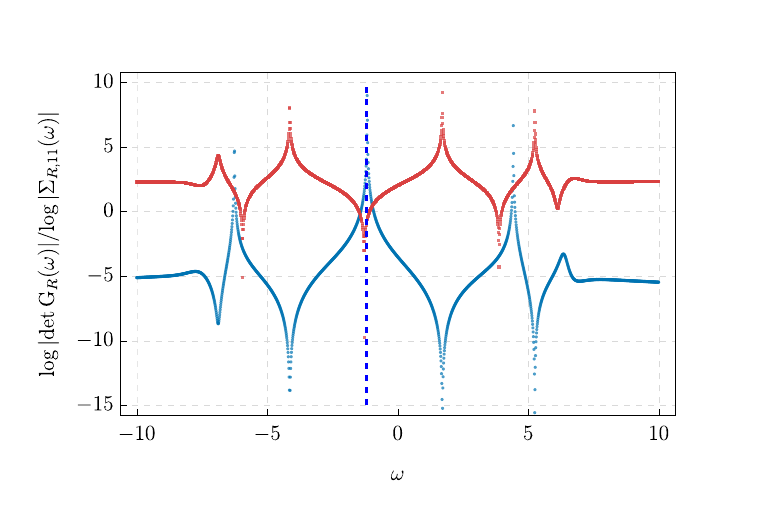}
  \caption{$\eta=5$}
  \label{fig:DetGSigmaMetal}
\end{subfigure}
\caption{Comparison of $\log\det \vert G_R(k=0,\omega) \vert$ (blue curves) and $\log\vert \Sigma_{R,11}(k=0,\omega) \vert$ (red curves) at $\eta=\pm5$. The dashed gray lines indicate the edges of the gap in the Mott phase, while the blue dashed line marks the location of the Fermi level in the metallic phase.}
\label{fig:GMottMetal}
\end{figure}

In addition, plots of $\vert \det G_R(\omega,k,\eta) \vert$ together with the real part of the self-energy at the same parameter values are presented in Fig.~\ref{fig:DetGSigmaMott}. In the case of $\eta=-5$, the self-energy exhibits a pole at the location of the zero of the Green's function, whereas in the other case the Green's function develops a pole near a zero of $\S_{R,11}$, in agreement with the large-$N$ argument presented in Sec.~\ref{sec:poleszeroessemiholo}.

\subsection{Crossover between Mott insulator and semi-holographic metal}

We now discuss the phenomenology of the crossover from the incoherent metallic phase close to $\eta=0$ to either a Mott insulating phase or a semi-holographic metallic phase within our semi-holographic model.
We focus on the $G_{R,11}$ component of the single-particle retarded Green's function, as the second diagonal component is related by the symmetry
\begin{align}\label{eq:symmetry}
G_{R,22}(\omega,k) = G_{R,11}(\omega,-k),
\end{align}
and can be obtained straightforwardly via the mapping $k \mapsto -k$.
\begin{figure}[t!]
  \centering  \includegraphics[width=0.8\linewidth]{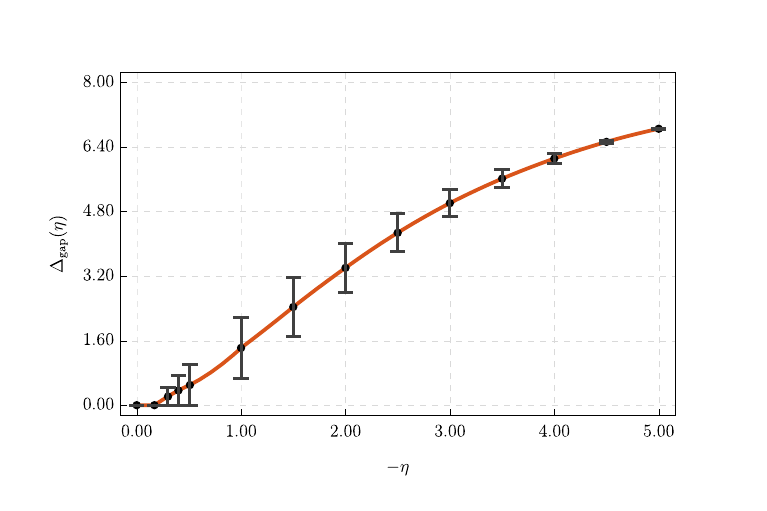}
  \caption{For negative values of $\eta$, the system is a Mott insulator and here we plot its gap size as a function of $\eta<0$. The gap opens at $\eta=-0.17$ for the cutoff $10^{-2}$ and at $\eta=-0.51$ for the $10^{-5}$ cutoff. The datapoints are determined from the average of the result for the gap for the two different cutoffs, while the error bars are given by the deviation from the average. For larger negative values of $\eta$ the two criteria converge towards the same value due to the spectral features becoming sharper.}
\label{fig:GapPlot}
\end{figure}
\begin{figure}
\begin{subfigure}{.5\textwidth}
  \centering
  \includegraphics[width=1\linewidth]{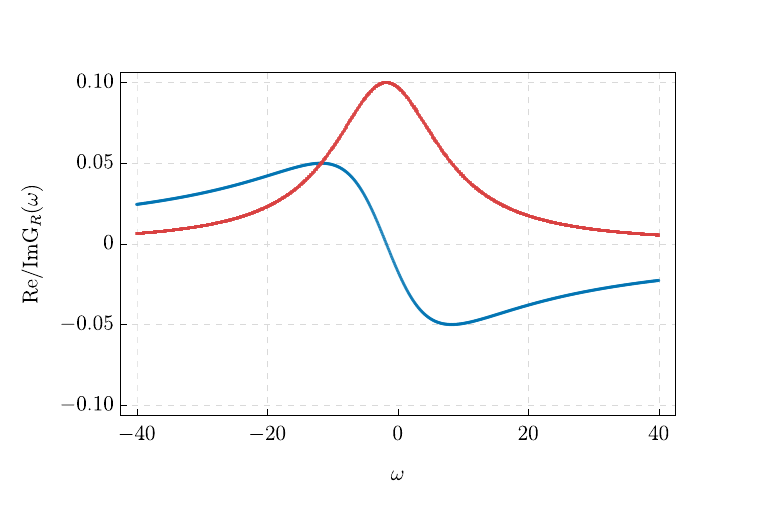}
  \caption{$\eta=0$}
  \label{fig:MottGap0}
\end{subfigure}
\begin{subfigure}{.5\textwidth}
  \centering
  \includegraphics[width=1\linewidth]{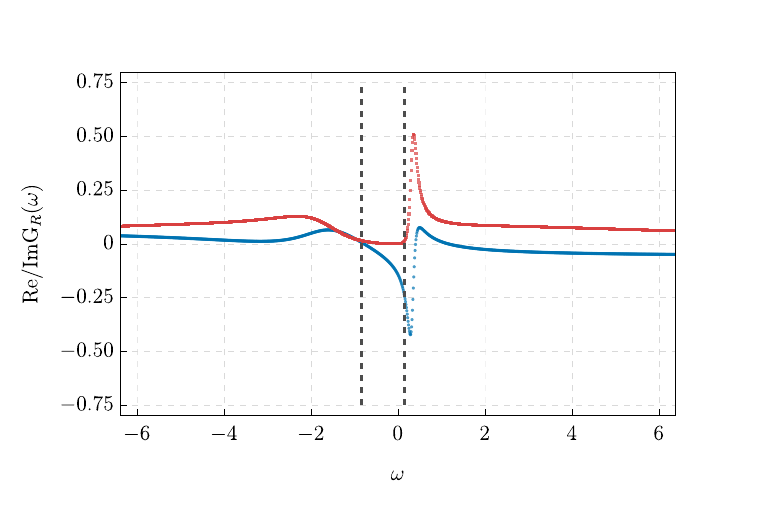}
  \caption{$\eta=-0.5$}
  \label{fig:MottGap05}
\end{subfigure}
\begin{subfigure}{.5\textwidth}
  \centering
  \includegraphics[width=1\linewidth]{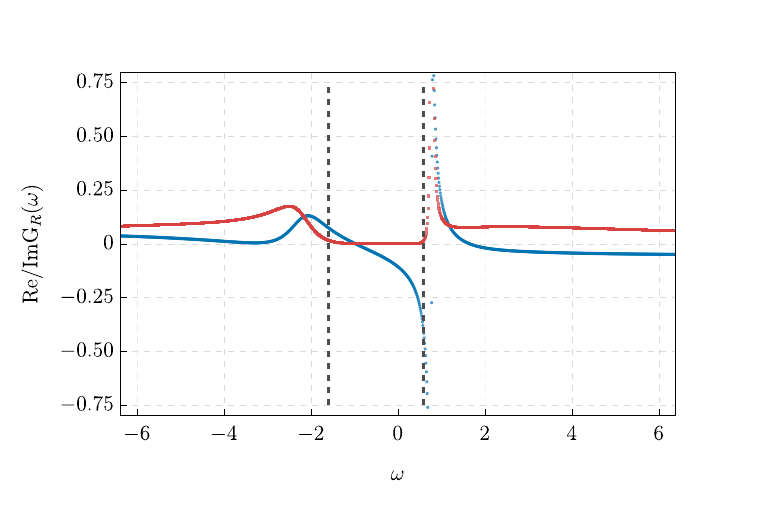}
  \caption{$\eta=-1$}
  \label{fig:MottGap1}
\end{subfigure}
\begin{subfigure}{.5\textwidth}
  \centering
  \includegraphics[width=1\linewidth]{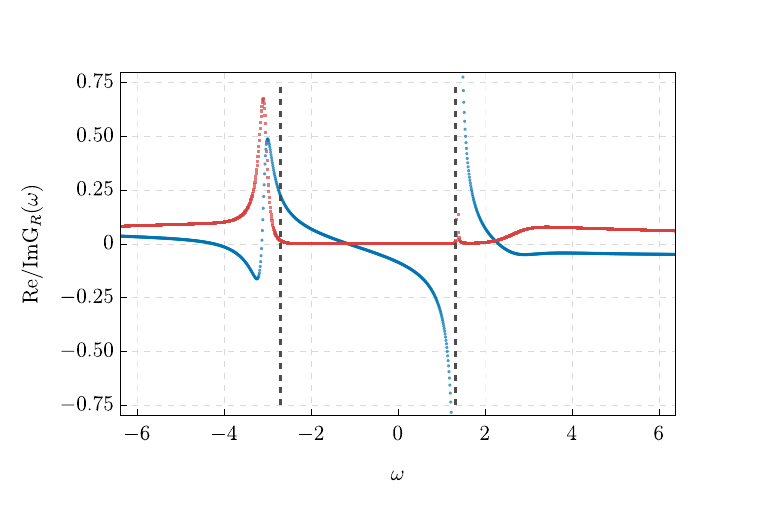}
  \caption{$\eta=-2$}
  \label{fig:MottGap2}
\end{subfigure}
\caption{Plots of the real (blue) and imaginary (red) part of $G_\mathrm{R}$ at $k=0$ over $\o$ for various negative values of the bulk coupling $\eta$ in increasing order. The gap, if it exists, is marked with the gray dashed lines. In the gapped phase there is a zero of the real part of the Green's function inside the gap as can be seen in Fig. \ref{fig:MottGap2}. At the location of the Hubbard bands one can see the typical linear, albeit very steep, increase behavior of the retarded Green's function. Increasing the bulk coupling results in moving not only the zero, but also shifting spectral weight into the once gapped region until the gap fully closes.}
\label{fig:GapClosing}
\end{figure}

Figure~\ref{fig:GapPlot} shows the behavior of the gap size
$\Delta_{\mathrm{gap}}$ as the control parameter $-\eta$ is increased to
positive values, for the two different cutoffs considered. The upper and lower values of the error bars correspond to the more lenient and stringent cutoff respectively. The respective datapoints are calculated from the average of the two values. In both cases we observe that $\Delta_{\mathrm{gap}}$ increases as $\eta$ becomes more negative. At larger negative values of $\eta$, as the broad spectral features in $\rho(\omega,k)$ become increasingly sharp, the two estimates of the gap begin to converge. For values of $\eta$ close to zero, the two-point function exhibits a very broad quasiparticle peak (see Fig.~\ref{fig:MottGap0}). As $\eta$ is further decreased, this quasiparticle feature splits into two distinct structures (see Figs.~\ref{fig:MottGap05}--\ref{fig:MottGap2}). The spectral weight removed from the Fermi level is redistributed as clearly visible in Fig.~\ref{fig:MottGap2}. Due to the presence of the in-gap Green's function zeros, these two high-frequency features can be interpreted as the Hubbard bands of a Mott insulator. One of the Hubbard bands is less pronounced in our results due to the nonzero bulk charge $q \neq 0$, which induces an $\omega$-asymmetry in the Green's function \cite{Ghorai:2024nxs}. Only at larger bulk couplings do both features become clearly visible; see, for instance, Fig.~\ref{fig:MottGap2}.

Our operational criterion for Mottness is the appearance of an isolated zero of $\mathrm{Det}\, G_R$ residing within the spectral gap. In Fig.~\ref{fig:DetGSigmaMott} we can already observe this behavior for a large negative $\eta$, however we can find similar results for other gapped states in e.g. Figs.~\ref{fig:MottGap1}--\ref{fig:MottGap2}. For our method of gap determination the zero does not always lie inside a gap at very low values of $\eta$, but it is very close to it, see e.g. Fig.~\ref{fig:MottGap05}. The gap itself opens up at either $\eta=-0.17$ or $-0.51$ for the two choices of the cutoff. In both cases this simply corresponds to the point when sufficient spectral weight (in accordance with the respective cutoff) is transferred into the emerging Hubbard bands. Note that it is the $10^{-2}$ cutoff in particular which aligns well with both sharp and broad spectral features, as can be seen in e.g. Fig. \ref{fig:GapClosing}. We remark that it is precisely the progressive sharpening of the collective excitation (namely, the narrowing of the peaks associated with $\mathcal{G}_R$) that opens the gap and places the zero within it. The semi-holographic setup thus suggests, that Mottness itself is caused by the emergent singular structure of these collective modes, which in our case appear as simple divergencies of a two-point function.
\begin{figure}
\begin{subfigure}{.5\textwidth}
  \centering
  \includegraphics[width=1\linewidth]{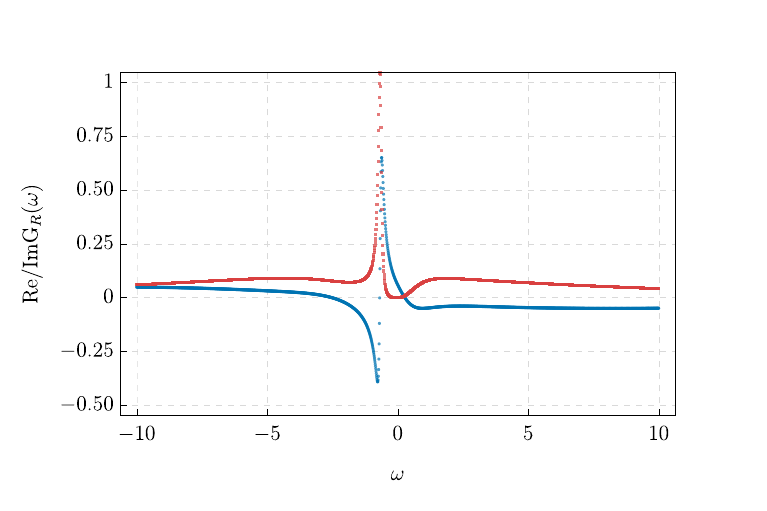}
  \caption{$\eta=0.5$}
  \label{fig:Metal0d5}
\end{subfigure}
\begin{subfigure}{.5\textwidth}
  \centering
  \includegraphics[width=1\linewidth]{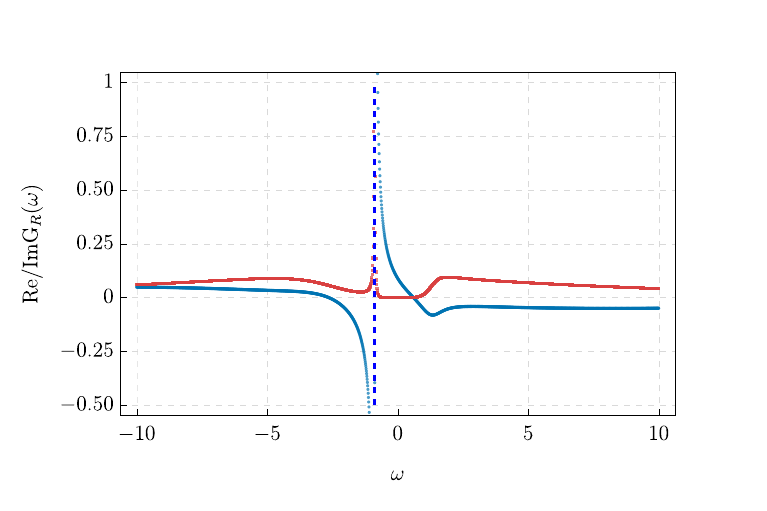}
  \caption{$\eta=1$}
  \label{fig:Metal1}
\end{subfigure}
\begin{subfigure}{.5\textwidth}
  \centering
  \includegraphics[width=1\linewidth]{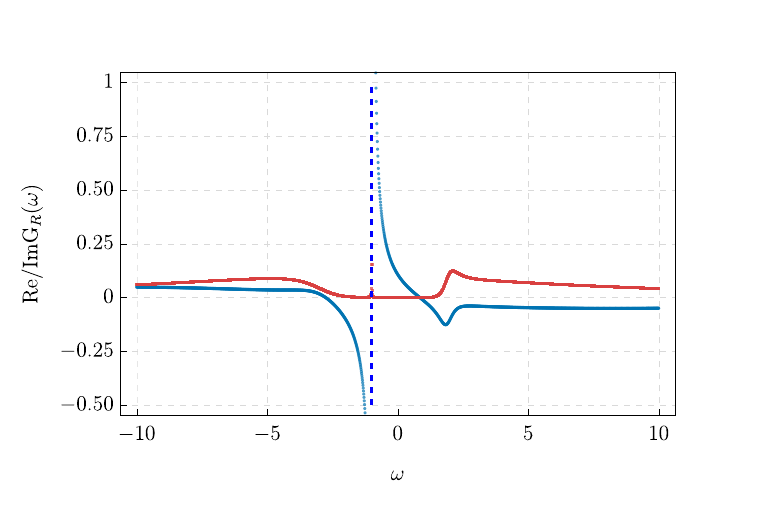}
  \caption{$\eta=1.5$}
  \label{fig:Metal1d5}
\end{subfigure}
\begin{subfigure}{.5\textwidth}
  \centering
  \includegraphics[width=1\linewidth]{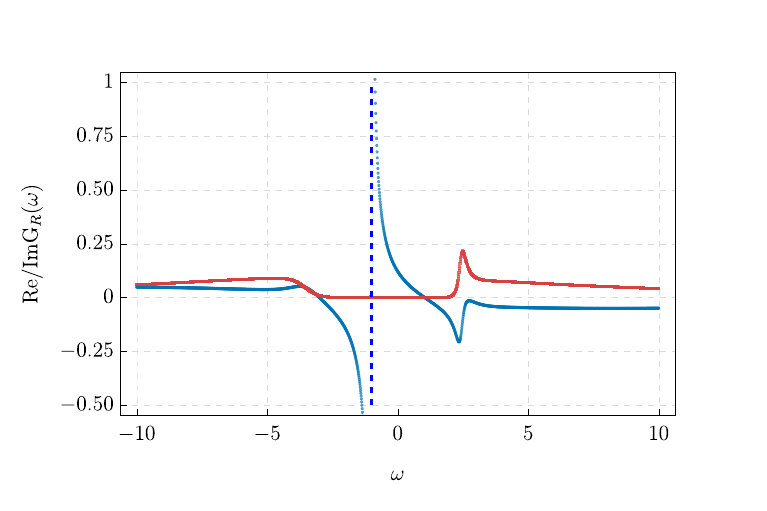}
  \caption{$\eta=2$}
  \label{fig:Metal2}
\end{subfigure}
\caption{Plots of the real (blue) and imaginary part (red) of $G_\mathrm{R}$ at $k=0$ over $\o$ for various positive values of the bulk coupling $\eta$ in increasing order. The Fermi levels are marked with a blue line. For increasing values of $\eta$ the original quasiparticle peak from Fig. \ref{fig:MottGap0} sharpens until turning into a pole, signaling a crossover from a incoherent metallic phase to a more coherent, although still strongly renormalized, semi-holographic metal. We note that in Fig.~\ref{fig:Metal2} the pole in the imaginary part is not visible due both to the blue curve and to its extreme sharpness.}
\label{fig:MetallicPlots}
\end{figure}

The crossover into the metallic regime for increasing positive values of $\eta$ is characterized by a progressive sharpening of the initially broad peak associated to the strong correlation. Representative spectra for this direction of parameter change are plotted in Fig.~\ref{fig:MetallicPlots}. Comparing the pole locations in Fig.~\ref{fig:MetallicPlots} with those in Fig.~\ref{fig:GapClosing} for matching $\vert\eta\vert$ reveals - in line with the discussion in the previous section - an approximate matching between the pole near the Fermi level for one sign of $\eta$ and the zero that appears for the opposite sign. Deep in the metallic phase, as show in Fig. \ref{fig:MetallicPlots}, we also see that spectral weight has been further transferred away from the peak at Fermi level into broadened peaks further away from it. These should be interpreted as the valence- and conduction band respectively. We note that even here the real part of the trace of the matrix residue at the pole, as shown by the blue vertical line in Fig.~\ref{fig:DetGSigmaMetal} at $\omega_p=-1.17$, is $\mathrm{Tr}\; \mathrm{Res}\, G_{R}(k=0,\omega_p=-1.17)\approx0.498$ with contributions that split equally between the components of the Green's function. A pole strength given by a residue of this magnitude indicates that we are looking at a coherent, although noticeably renormalized, metal.

\section{Discussion}
\label{sec:discuss}

We have proposed a semi-holographic model exhibiting a Mott-insulating phase. A fundamental fermionic field interacts with a strongly coupled sector that generates a self-energy for the fundamental fermion's Green's function $G_{R}$. The latter, due to the semi-holographic construction, satisfies single-particle sum rules and develops a poles–zeros duality at low frequency and momenta. Mott-insulating phases are characterized by the presence of in-gap {Green's function} zeros. In this respect, the poles–zeros duality plays a crucial role in interpreting the model as a candidate description of Mott physics. In the semi-holographic picture, the zeros of the Mott-insulating phase are caused by many-body excitations of the strongly interacting sector.

{The duality of poles and zeros we observe in our setup is a consequence of the large-$N$ limit inherent to our holographic description. In this limit, the self-energy is dominated by the two-point function of the composite operator $\mathcal{O}$, which develops either simple poles in the Mott insulating phase, or simple zeros in the phase we identify as a semi-holographic metal.}

{We performed a numerical analysis focused on the Green's functions $G_{R}$ of the fundamental field at zero temperature. We found that the approximate duality is already very accurate at $N\sim \vert\alpha_f g^2\vert=10$. We have then studied the behavior of poles and zeros as a function of the model parameter $\eta$, demonstrating that it drives the system from a semi-holographic metal to a Mott-insulating phase. This manifests itself respectively as poles or zeros in the determinant of the Green's function at low frequencies. In the Mott phase, we find that the zeros lie inside the spectral gap, demonstrating that the semi-holographic construction properly reproduces the known phenomenology of Mott insulating systems.}

In the condensed-matter context, the duality of poles and zeros in Green’s functions \cite{wagner2023,lehmannPRL2025} has been observed in many-body solutions of Hubbard models, where it emerges from divergences of the self-energy in the Mott insulating phase. These singularities exhibit a momentum dispersion that closely resembles the free energy-momentum relation of the Fermi-liquid phase at weak coupling \cite{wagner2023,lehmannPRL2025,setty_electronic,setty_symmetry_2023,Blason_2023,gleis_PRX_2024,pangburn_PRB_2025}. An important difference with respect to the present analysis is that, in this many-particle description of Hubbard Hamiltonians, poles and zeros appear on opposite sides of a thermodynamic phase transition separating the Fermi-liquid metal from the Mott insulator, so the relation between the momentum dispersion of the zeros and the free band structure is not a simple mapping. By contrast, our semi-holographic construction makes the emergence of poles-zeros duality more transparent, since it arises from a switching of the strongly coupled bath CFT from standard to alternative quantization and vice versa. In this framework, the change of sign of a control parameter brings the system from low-frequency poles to in-gap zeros, so the phase transition is effectively replaced by an {\it inversion} of the large-$N$ self-energy $\Sigma_R$ associated with the strongly interacting holographic background. Close to or at the quantum critical point, the semi-holographic model thus provides a natural description of the low-energy dynamics in terms of the conformal scaling data of the composite operator $\mathcal{O}$. More broadly, this suggests that a large class of systems exhibiting a duality between poles and zeros in their conducting and Mott-insulating states can, after integrating out a large number of degrees of freedom, be understood as the result of strongly interacting operator dynamics whose gravity dual description admits an exchange between standard and alternative quantization schemes. Describing poles and zeros in this way may therefore offer a useful perspective on their conjectured common nature as excitations \cite{fabrizio_emergent_2022,fabrizio_PRL_2023,Wagner_2024}.

Our work paves the way for several future directions of study. In the present paper, we have focused on the effects of the bulk scalar term with coupling parameter $\eta$ (see Eq.~(\ref{eq:completeholoaction})), which gives rise to the exact poles–zeros duality of the self-energy. In fact, there exist further couplings (see e.g. \cite{Ghorai:2024nxs}) that would also entail a poles–zeros duality of the self-energy. It would be interesting to address these other couplings within the semi-holographic setup and to study how the corresponding phenomenology would differ, especially in relation to Mott physics. Relatedly, it will be interesting to further investigate the origin of the standard-to-alternative quantization transition behind the poles-zeros duality by calculating the infrared exponents of the composite fermionic operator.

A further  interesting question concerning Mott-insulating phases is whether they may also exhibit non-trivial topological properties. A current debate in the literature concerns the meaning for Hall responses of topological invariants defined through the single-particle Green's function \cite{markovARXIV, Blason_2023, peralta2023, Lessnich2024}. The present semi-holographic construction can be extended to the study of current-current correlation functions and conductivities. This will be the object of a future study aimed at shedding light on the relation between the non-trivial winding in momentum space of Green's function zeros and physical observables of strongly interacting phases. Our semi-holographic approach in particular allows the analysis of the contributions of both poles and zeros of the fermionic Green's function to the current-current correlator entering the optical conductivity, as well as the study of the conductivity's behavior under the poles-zeros duality. 

Finally, holography in principle also allows the observation of a phase transition from the semi-holographic metallic to the Mott insulating phase. In this work, this transition is not distinguishable from the change in behavior of the composite fermionic two-point function due to the large-$N$ limit preventing the composite operator from condensing and backreacting onto the background geometry. This can be overcome, following \cite{vcubrovic2009string,vcubrovic2011constructing,vcubrovic2011spectral,gubankova2011holographic,medvedyeva2013quantum}, by backreacting the fermionic determinant in the bulk path integral onto the geometry, or by Tolman-Oppenheimer-Volkoff  fluid approximations as in \cite{de2010holographic}. This will in particular allow to further characterize the infrared fixed point at the phase transition from holography. 
Last but not least, the results of the present paper invite to revisit studies of sum rules in holography in connection with universal behaviour  \cite{Erdmenger:2012ik,Erdmenger:2015qqa}, also including broken translational symmetry in presence of a lattice. 
We plan to come back to all of these questions in future work.

\section*{\large Acknowledgments}

We thank Jani Kastikainen and Souvik Banerjee for helpful comments. A.C.,
J.E., T.K., R.M., F.P. and G.S. acknowledge the support of the German Research Foundation (DFG) through the Collaborative Research Center ToCoTronics, Project-ID 258499086 — SFB 1170, as well as Germany's Excellence Strategy through the W{\"u}rzburg-Dresden Cluster of Excellence on Complexity, Topology and Dynamics in Quantum Matter - ctd.qmat (EXC 2147, Project-ID 390858490). R.M.  acknowledges hospitality from the Shanghai Institute for Mathematics and Interdisciplinary Sciences (SIMIS) and associated travel support under STCSM Grant 25HB2701900.

\bibliographystyle{utphys}
\bibliography{bibliography}

\end{document}